\documentclass[onecolumn]{revtex4-1}

\usepackage{graphicx}
\usepackage{dcolumn}
\usepackage{bm}
\usepackage{amsmath}
\usepackage{mathtools}
\DeclarePairedDelimiter\bra{\langle}{\rvert}
\DeclarePairedDelimiter\ket{\lvert}{\rangle}
\DeclarePairedDelimiterX\braket[2]{\langle}{\rangle}{#1 \delimsize\vert #2}

\newcommand{\tens}[1]{%
	\mathbin{\mathop{\otimes}\limits_{#1}}%
}

\let\oldAA\AA
\renewcommand{\AA}{\text{\normalfont\oldAA}}

\usepackage{amsmath}
\usepackage{subfigure}
\usepackage{hyperref}
\usepackage[normalem]{ulem}
\hypersetup{
    colorlinks,
    citecolor=blue,
    filecolor=black,
    linkcolor=blue,
    urlcolor=blue
}
\usepackage{epstopdf}
 \usepackage{relsize}

\newcommand*{\rom}[1]{\expandafter\@slowromancap\romannumeral #1@}
\title{Report}

\setcitestyle{square}

\begin{document}


\title{Density matrix to quantum master equation (QME) model  for arrays of Coulomb coupled quantum dots in the   sequential tunneling regime  }

\author{Aniket Singha}
\affiliation{%
Department of Electronics and Electrical Communication Engineering,\\
Indian Institute of Technology Kharagpur, Kharagpur-721302, India\\
}%





\begin{abstract}
Coulomb coupled quantum dot arrays with staircase ground state configuration have been proposed in literature for enhancing heat-harvesting and refrigeration performance \cite{coulomb_TE7,coulomb_TE1,coulomb_TE2,coulomb_TE3,coulomb_TE4,coulomb_TE5,coulomb_TE6,aniket_cool1}. Due to their  mutual Coulomb interaction, a performance analysis of such systems remains complicated and necessitates consideration of microscopic physics using density matrix formulation. However the path of transport analysis starting from the system Hamiltonian  to density matrix formulation is complicated and lacks the simplicity and intuitive aspect of sequential electron transport conveyed by the  quantum master equation (QME) approach. In this paper, starting from the system Hamiltonian and employing the density matrix formulation, I derive the QME of a system of three quantum dots, two of which are electro-statically coupled. The framework elaborated in this paper can  be further extended  to derive QME of systems with  higher number of Coulomb coupled quantum dots. Hence, the formulation developed in this paper can pave the way towards an intuitive analysis of transport physics for an array of Coulomb coupled quantum dots in the sequential tunneling regime.
\end{abstract}
\maketitle

\begin{figure}
	\centering
	\includegraphics[scale=.12]{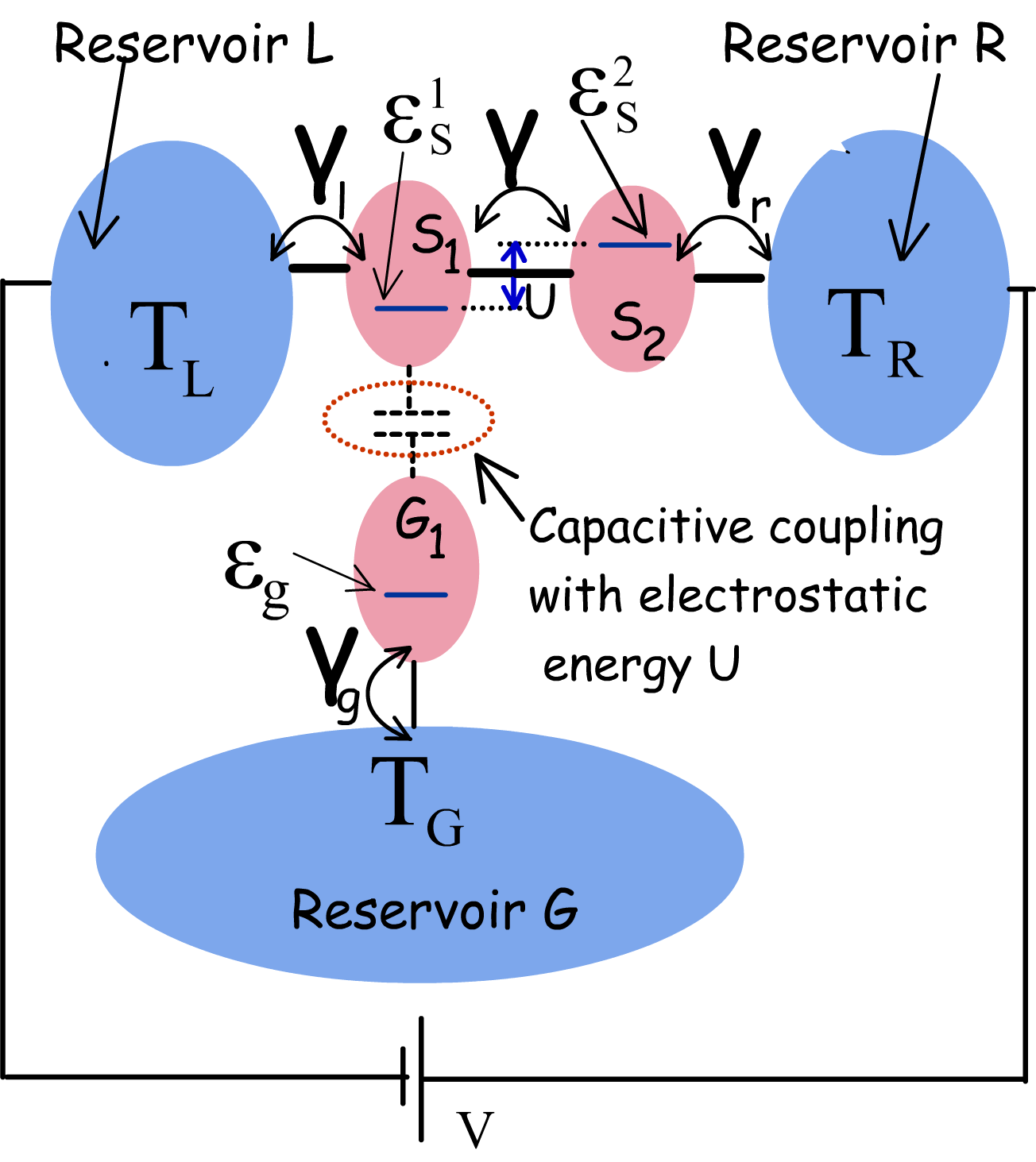}
	\caption{ Schematic of a system of coupled quantum dots $S_1$, $S_2$ and $G_1$ The dot $S_1$ and $S_2$ are electrically connected to the reservoirs $L$ and $R$ respectively, while $G_1$ is electrically connected to the reservoir $G$. $S_1$ and $S_2$ are tunnel coupled while $G_1$ and $S_1$ are capacitively coupled with a mutual charging energy $U$}
	\label{fig:Fig1}
\end{figure}

 Recently with the progress of fabrication technology, a lot of effort has been geared towards nanoscale solid state quantum dot devices which, due to their discrete energy spectrum, form ideal beds of quantum computation, heat harvesting and refrigeration, etc. Due to their small size, quantum dots that are separated in space often exhibit capacitive charge coupling which offers another degree of freedom to manipulate charge, energy and spin. This phenomenon of electrostatic or charge-based coupling between spatially separated quantum dots is known as Coulomb coupling, which gives rise to a well known phenomena known as \textit{Coulomb blockade.} Quantum dots that are spatially separated, may be bridged to obtain strong electrostatic coupling between them \cite{cap_coup_1,cap_coup_2}. In addition, the bridge may be fabricated between two desired  dots to radically increase their mutual electrostatic coupling, without affecting the other dots \cite{cap_coup_1,cap_coup_2,cap_coup3}. The effect of Coulomb coupling on the current spectra  as well as methods to enhance Coulomb coupling between quantum dots has been well explored via experiments \cite{cap_coup3,cap_coup_2,cap_coup_1}. However, a theoretical analysis of such Coulomb coupled dots is complicated and requires a full analysis of the microscopic physics starting from the system Hamiltonian. This is particularly true when  a few or all of the dots in the system are Coulomb coupled to one or more adjacent dots. In addition, the analysis approach of such set-ups, starting from the system Hamiltonian, often masks the intuitive aspect of  sequential transport physics, which is generally beneficial to the experimental community to further refine device characteristics. \\
\indent In this paper, starting from the microscopic physics, I methodically derive the quantum master equations (QME) of  Coulomb-coupled dot arrays in the sequential tunneling limit. In addition to being a simpler framework, the QME approach \cite{master_eq_1,master_eq_2,master_eq_3,master_eq_4,master_eq_5,master_eq_6}  bears the intuitive aspect of sequential electron transport and have been extensively used in literature to analyze the properties of single electron transistors and quantum dots in the Coulomb blockade regime. However, to the best of my knowledge, such approach has not yet been used to analyze arrays of quantum dots in which one or more dots may be Coulomb coupled with some others. Here, starting from the system Hamiltonian and the density matrix formulation, I derive the quantum master equations of the Coulomb coupled system demonstrated in Fig.~\ref{fig:Fig1}. Such type of Coulomb coupled systems have already been proposed for the optimal non-local refrigeration \cite{coulomb_TE7}. The system consists of three dots $S_1,~S_2$, and $G_1$ which  are electrically coupled to the reservoirs $L$, $R$ and $G$ respectively. $S_1$ and $S_2$ are tunnel coupled to each other, while $G_1$ is capacitively coupled to $S_1$. The ground states of $S_1$ and $S_2$ form a stair-case configuration with $\varepsilon_s^2\approx \varepsilon_s^1+U$.\\
\indent To derive the quantum master equations of the system, I start from the device Hamiltonian.  The increase in total  total electrostatic energy $E_C$ of the system consisting of three dots, due to fluctuations from the reservoirs, can be given by: 
\begin{widetext}
\small
\begin{eqnarray}
E_C(n_{S_{1}},n_{G_{1}},n_{S_{2}})=\sum_{x \in (S_{1},G_{1},S_{2})}E^{self}_{x}\left(n_{x}^{tot}-n_x^{eq}\right)^2  +\sum_{(x_{1},x_{2})\in(S_{1},G_{1},S_{2})}^{x_1 \neq x_2} E^m_{x_1,x_2}\left(n_{x1}^{tot}-n_{x1}^{eq}\right)\left(n_{x2}^{tot}-n_{x2}^{eq}\right) \nonumber
\end{eqnarray}
\end{widetext}
\normalsize
where $n_x^{tot}$ is the total electron number,   and $E^{self}_x=\frac{q^2}{C^{self}_{x}}$ is the electrostatic energy due to self-capacitance $C^{self}_{x}$  of    quantum dot `$x$' with its surrounding terminals. $E^m_{x_1,x_2}$ is the electrostatic energy arising out of interdot  Coulomb interaction between two different quantum dots that are separated in space. $n_x^{eq}$ is the electron number at system equilibrium at $0K$ and is to be determined by the minimum possible electrostatic energy of the system. $n_x=n_x^{tot}-n_x^{eq}$ is the number of electrons in the ground state of the dot $x$. The electron number in the ground state of the quantum dots may fluctuate at finite temperature due to fluctuations from the reservoirs.  Here, a minimal physics based model is used to derive the rate equations. I assume that the electrostatic energy due self-capacitance is much greater than than the average thermal voltage $kT/q$ or the applied bias voltage $V$, that is $E^{self}_x=\frac{q^2}{C^{self}_{x}}>> (kT,~qV)$, such that electron occupation probability or transfer rate via the Coulomb blocked energy level, due to self-capacitance, is negligibly small.  The analysis  of the entire system of dots may hence be approximated by limiting the maximum number of electrons in each  dot to one. Thus the analysis of the entire system may be limited to eight multi-electron  levels, which I denote by the electron occupation number in the ground state of each quantum dot. Hence, a possible state of interest in the system may be denoted  as $\ket{n_{S_1},n_{G_1},n_{S_2}}=\ket{n_{S_1}}\tens{} \ket{n_{G_1}} \tens{} \ket{n_{S_2}}$, where $(n_{S_1},n_{G_1},n_{S_2})\in (0,1)$. To proceed further from here, with a slight abuse of notation, I simply denote the eight multi-electron states as $\ket{0,0,0}\rightarrow \ket{0}$, $\ket{0,0,1}\rightarrow \ket{1}$, $\ket{0,1,0}\rightarrow \ket{2}$, $\ket{0,1,1}\rightarrow \ket{3}$, $\ket{1,0,0}\rightarrow \ket{4}$, $\ket{1,0,1}\rightarrow \ket{5}$, $\ket{1,1,0}\rightarrow \ket{6}$, and 
$\ket{1,1,1}\rightarrow \ket{7}$.
The Hamiltonian of the  system consisting of these three quantum dots without any reservoir coupling  may be written as:
\begin{eqnarray}
H=\sum_{\beta}\epsilon_{\beta}\ket{\beta}\bra{\beta}+t\{\ket{3}\bra{6}+\ket{1}\bra{4}\} +U \{\ket{6}\bra{6}+\ket{7}\bra{7}\}+h.c.,
\end{eqnarray}
where $U=E^m_{S_1,G_1}$ is the Coulomb coupling energy between the dots $S_1$ and $G_1$ in Fig.~\ref{fig:Fig1} and $t$ is the electron hopping amplitude between the adjacent dots $S_1$ and $S_2$.  Under the assumption of weak reservoir to system coupling and  small hopping amplitude  $t$, the temporal dynamics of the system density matrix can be evaluated by  the partial trace over the  density matrix of the entire set-up  of the reservoirs and the dots \cite{master_eq_1,master_eq_2,master_eq_3,master_eq_4,master_eq_5,master_eq_6}. Taking the partial trace of the combined density matrix over the reservoir states, the diagonal and the non-diagonal elements of the density matrix $\rho$ of the system   of quantum dots may be given as a set of modified Liouville euqations \cite{master_eq_1,master_eq_2,master_eq_3,master_eq_4,master_eq_5,master_eq_6}:
 \begin{eqnarray}
\frac{\partial \rho_{\eta \eta}}{\partial t}=-i[H,\rho]_{\eta \eta}-\sum_{\nu} \Gamma_{\eta \nu}\rho_{\eta \eta}+\sum_{\delta}\Gamma_{\delta \eta }\rho_{\delta \delta} 
\frac{\partial \rho_{\eta \beta}}{\partial t}=-i[H,\rho]_{\eta \beta}-\frac{1}{2}\sum_{\nu} \Big(\Gamma_{\eta \nu}+\Gamma_{\beta \nu }\Big) \rho_{\eta \beta},
\label{eq:time_derivate}
\end{eqnarray}
where $[A,B]$ denotes the commutator of  $A$ and $B$ and $\rho_{\eta \beta}=\bra{\eta}\rho \ket{\beta}$. The elements $\rho_{\eta \eta}$ and $\rho_{\eta \beta}$ in the above equation denote any diagonal and non-diagonal element of the system density matrix respectively. The parameters $\Gamma_{ij}$ take into account the transition between system   states  due to electronic tunneling between the system and the reservoirs and are only non-zero when the system can transit from state $\ket{i}$ to $\ket{j}$ (or vice-versa) due to tunneling of electrons in and out of the system from the reservoir. In our derivation, assuming  a statistical quasi-equilibrium distribution of electrons inside the reservoirs, we can express $\Gamma_{i j}$  as:   
\begin{eqnarray}
\Gamma_{i j }={\gamma_c}f_{\lambda}({\epsilon_{i}}-\epsilon_{j}), 
\label{eq:c4}
\end{eqnarray}
  $f_{\lambda}(\epsilon)$ being  occupancy probability of  the corresponding reservoir $\lambda$  at energy $\epsilon$ and $\epsilon_{i (j)}$ is the total electronic energy of the system in the state $i~(j)$.  \\
To derive the quantum master equations for the entire system, it is essential to derive the inter-dot tunneling rates.  For the particular  system schematic demonstrated  in Fig.~\ref{fig:Fig1}, interdot tunneling changes the system states as: $\ket{4} \longleftrightarrow \ket{1}$ and  $\ket{3}\longleftrightarrow  \ket{6}$. Taking the time derivative of the density matrix to be zero in steady state, I use the second equation of \eqref{eq:time_derivate}, to  get, 
\begin{eqnarray}
\rho_{4,1}=\rho^{*}_{1,4}=\frac{\rho_{4,4}-\rho_{1,1}}{\epsilon_{4}-\epsilon_{1}-i\frac{\Omega_{4,1}}{2}}
\label{eq:c5}
\\
\rho_{6,3}=\rho^{*}_{3,6}=\frac{\rho_{6,6}-\rho_{3,3}}{\epsilon_{6}-\epsilon_{3}-i\frac{\Omega_{6,3}}{2}},
\label{eq:c6}
\end{eqnarray}
where $\Omega_{i,j}$ is the combination  of all the reservoir-to-system tunneling events (or vice-versa) leading to the decay of the states ${i}$ and ${j}$. For the system under consideration, $\Omega_{4,1}$ and $\Omega_{6,3}$ can be given by:
\begin{gather}
\Omega_{4,1}=\Gamma_{{4},{0}}+\Gamma_{{4},{6}}+\Gamma_{{4},{5}}+\Gamma_{{1},{0}}+\Gamma_{{1},{6}}+\Gamma_{{1},{3}} \nonumber \\
\Omega_{6,3}=\Gamma_{{6},{4}}+\Gamma_{{6},{2}}+\Gamma_{{6},{7}}+\Gamma_{{3},{1}}+\Gamma_{{3},{2}}+\Gamma_{{3},{7}} 
\end{gather}
The time derivative of diagonal density matrix elements $\rho_{6,6}$ and $\rho_{3,3}$  can be written as (using the first equation of \ref{eq:time_derivate}):
\begin{align}
\dot{\rho}_{6,6}=&it(\rho_{6,3}-\rho_{3,6})-\left(\Gamma_{{6},{4}}+\Gamma_{{6},{2}}+\Gamma_{{6},{7}}\right)\rho_{6,6}+\Gamma_{{4},{6}}\rho_{4,4}+\Gamma_{{2},{6}}\rho_{2,2}+\Gamma_{{7},{6}}\rho_{7,7} \nonumber \\
\dot{\rho}_{4,4}=&it(\rho_{4,1}-\rho_{1,4})-\left(\Gamma_{{4},{0}}+\Gamma_{{4},{6}}+\Gamma_{{4},{5}}\right)\rho_{4,4}+\Gamma_{{0},{4}}\rho_{0,0}+\Gamma_{{6},{4}}\rho_{6,6}+\Gamma_{{5},{4}}\rho_{5,5} 
\label{eq:b8}
\end{align}
I next substitute, in Eq.~\eqref{eq:b8}, the expressions for $\rho_{6,3},~\rho_{3,6},~\rho_{4,1}$ and $\rho_{1,4}$   from Eq.~\eqref{eq:c5}, to get the time evolution of the  density matrix elements of $\rho_{6,6}$ and $\rho_{4,4}$  as:
\begin{eqnarray}
\dot{p_6}=\dot{\rho}_{6,6}=\sum_{\alpha}\left(  -\Gamma_{{6},{\alpha}}p_6+\Gamma_{{\alpha},{6}}p_{\alpha} \right) 
-\tau_{{6},{3}}p_6+\tau_{{3},{6}}p_3 \nonumber \\
\dot{p_4}=\dot{\rho}_{4,4}=\sum_{\alpha}\left(  -\Gamma_{{4},{\alpha}}p_3+\Gamma_{{\alpha},{4}}p_{\alpha} \right) 
-\tau_{{4},{1}}p_4+\tau_{{1},{4}}p_1, \nonumber
\end{eqnarray}
where $p_{\eta}=\rho_{\eta,\eta}$ and 
\begin{gather}
\tau_{{6},{3}}=\tau_{{3},{6}}=t^2\frac{\Omega_{6,3}}{(\epsilon _6-\epsilon _3)^2+\frac{\Omega _{6,3}^2}{4}}\nonumber \\
\tau_{{4},{1}}=\tau_{{1},{4}}=t^2\frac{\Omega_{4,1}}{(\epsilon _4-\epsilon _1)^2+\frac{\Omega _{4,1}^2}{4}} 
\label{eq:tun_rate}
\end{gather}
\begin{figure*}
	\vspace{-.8cm}\includegraphics[width=1.05\textwidth]{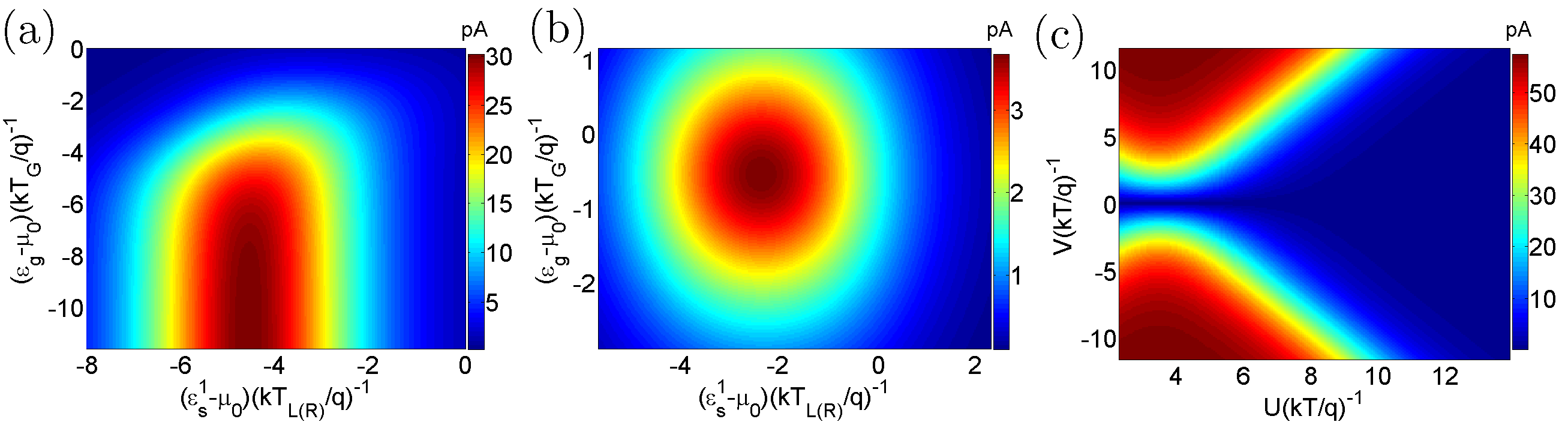}
\caption{ The characteristics of the set-up as predicted by the proposed QME. Colour plot for (a) current for a given voltage bias.  $T_R=T_L=T_G=T=5K,~U=2meV ~(\approx 4.63\frac{kT}{q})$ a voltage bias $V=1mV ~(\approx ~2.3\frac{kT}{q})$ (b) Short-circuited current for a given temperature bias.  $T_R=T_L=5K,~T_G=10K$ and $~U=2meV ~(\approx 3.1\frac{kT}{q})$ (c) Variation in current magnitude with applied voltage $V$ and Coulomb coupling energy $U$. $T_R=T_L=T_G=T=5K$ and $\varepsilon_s^1-\mu_0=-1.5meV~(\approx -3.5\frac{kT}{q})$ and $\varepsilon_g-\mu_0=-2meV~(\approx -4.63\frac{kT}{q})$.  $\mu_0$ is the equilibrium Fermi energy of the entire system and $T=\frac{T_{L(R)}+T_G}{2}$ is the average temperature of the reservoirs $L(R)$ and $G$.}
	\label{fig:Fig2}
\end{figure*}
In the set of Eqns.~\eqref{eq:tun_rate}, $\tau_{{4},{1}}$ and $\tau_{{6},{3}}$ correspond to the interdot tunneling rates when the number of electrons in dot $G_1$ is $0$ and $1$ respectively. When $U>>|\Omega _{4,1}|$, by an appropriate choice of $\epsilon_4-\epsilon_1$ and $\epsilon_6-\epsilon_3$, such that, $\epsilon_6=\varepsilon_g+\varepsilon_s^1+U=\varepsilon_g+\varepsilon_s^2=\epsilon_3$, that is by making $\varepsilon_s^2=\varepsilon_s^1+U$, we may arrive at a condition where $\tau_{{6},{3}}>>\tau_{{4},{1}}$. Such a condition implies that the tunneling probability between the dots is negligible in the absence of an electron in $G_1$, which is the combined impact of the capacitive coupling between $S_1-G_1$ and staircase ground-state configuration of $S_1-S_2$.\\
\indent Next, I proceed towards deriving the QME  of  the system demonstrated in Fig.~\ref{fig:Fig1}. Since, the electronic transport and ground states in $S_1$ and $G_1$ are mutually coupled, I  treat the pair of dots $S_1$ and $G_1$ as a  sub-system ($\varsigma_1$), $S_2$ being the complementary sub-system ($\varsigma_2$) of the entire system consisting of three dots. I assume that  $U>>|\Omega _{4,1}|$ and $\varepsilon_s^2=\varepsilon_s^1+U$, such that $\tau_{{4},{1}}<<\tau_{{6},{3}}$. For all practical phenomena relating to electron transport, it can hence be assumed that $\tau_{{4},{1}}\approx 0$. The state probability   $\varsigma_1$ is denoted by $P_{i,j}^{\varsigma_1}$,  $i$ and $j$ being the number of electrons in the dot $S_1$ and $G_1$ respectively. $P_k^{\varsigma_2}$, on the other hand, denotes the probability of occupancy of the dot $S_2$ in the sub-system $\varsigma_2$. Note that breaking down the entire system into two sub-system in this way is possible only in the limit of weak tunnel and Coulomb coupling between the two sub-systems, as such the state of one sub-system remains unaffected by a change in   state of the other sub-system. In such a limit, the diagonal elements of the density matrix can be written as:  $\rho_{0,0}=P^{\varsigma_1}_{0,0}P^{\varsigma_2}_{0},~\rho_{1,1}=P^{\varsigma_1}_{0,0}P^{\varsigma_2}_{1},~\rho_{2,2}=P^{\varsigma_1}_{0,1}P^{\varsigma_2}_{0},~\rho_{3,3}=P^{\varsigma_1}_{0,1}P^{\varsigma_2}_{1},~\rho_{4,4}=P^{\varsigma_1}_{1,0}P^{\varsigma_2}_{0},~\rho_{5,5}=P^{\varsigma_1}_{1,0}P^{\varsigma_2}_{1},~\rho_{6,6}=P^{\varsigma_1}_{1,1}P^{\varsigma_2}_{0},~\rho_{7,7}=P^{\varsigma_1}_{1,1}P^{\varsigma_2}_{1}$ The QME for the sub-system $\varsigma_1$ and $\varsigma_2$ can, hence, be derived by expressing the sub-system state probabilities   as the sum of two or more diagonal elements of the density matrix:\small
\begin{widetext}
	\begin{align}
	\frac{d}{dt}(P_{0,0}^{\varsigma_1})=\frac{d}{dt}\left( \rho_{0,0}+\rho_{1,1}\right)=&\gamma_c \times \left\{-P_{0,0}^{\varsigma_1}\{f_L(\varepsilon_s^1)+f_G(\varepsilon_g)\}+P_{0,1}^{\varsigma_1}\{1-f_G(\varepsilon_g)\}+P_{1,0}^{\varsigma_1}\{1-f_L(\varepsilon_s^1)\}\right\}\nonumber \\
	\frac{d}{dt}(P_{1,0}^{\varsigma_1})=\frac{d}{dt}\left(\rho_{5,5}+\rho_{4,4}\right)=&\gamma_c \times \left\{-P_{1,0}^{\varsigma_1}\left\{1-f_L(\varepsilon_{s}^1)+f_G(\varepsilon_g+U)\right\}+P_{1,1}^{\varsigma_1}\left\{1-f_G(\varepsilon_g+U)\right\}+P_{0,0}^{\varsigma_1}f_L(\varepsilon_s^1)\right\} \nonumber \\
	\frac{d}{dt}(P_{0,1}^{\varsigma_1})=\frac{d}{dt}\left(\rho_{2,2}+\rho_{3,3}\right)=&\gamma_c \times \left\{-P_{0,1}^{\varsigma_1}\left\{1-f_g(\varepsilon_{g}^1)+f_L(\varepsilon_s^1+U)+\frac{\gamma}{\gamma_c}P^{\varsigma_2}_1\right\}\right\} \nonumber \\& +\gamma_c \left\{P_{0,0}^{\varsigma_1}f_G(\varepsilon_g)+P_{1,1}^{\varsigma_1}\left\{1-f_L(\varepsilon_s^1+U)+\frac{\gamma}{\gamma_c}P^{\varsigma_2}_{0}\right\}\right\} \nonumber \\
	\frac{d}{dt}(P_{1,1}^{\varsigma_1})=\frac{d}{dt}\left(\rho_{7,7}+\rho_{6,6}\right)=&\gamma_c \times \left\{-P_{1,1}^{\varsigma_1}\left\{[1-f_g(\varepsilon_{g}^1+U)]+[1-f_L(\varepsilon_s^1+U)]+\frac{\gamma}{\gamma_C}P^{\varsigma_2}_0\right\}\right\} \nonumber \\ &+\gamma_c \left\{P_{1,0}^{\varsigma_1}f_G(\varepsilon_g+U) +P_{0,1}^{\varsigma_1}\left\{f_L(\varepsilon_s^1+U)+\frac{\gamma}{\gamma_c}P^{\varsigma_2}_{1}\right\}\right\} 
	\label{eq:first_sys}
	\end{align} 
	\begin{align}
	&\frac{d}{dt}(P_{0}^{\varsigma_2})=\frac{d}{dt}\left( \rho_{6,6}+\rho_{4,4}+\rho_{2,2}+\rho_{0,0}\right)=\gamma_c \times \left\{-P_{0}^{\varsigma_2}\{f_R(\varepsilon_s^2)+\frac{\gamma}{\gamma_c}P_{1,1}^{\varsigma_1}\}+P_1^{\varsigma_2}\{1-f_R(\varepsilon_{s}^2)+\frac{\gamma}{\gamma_c}P^{\varsigma_1}_{0,1}\}\right\}\nonumber \\
	&\frac{d}{dt}(P_{1}^{\varsigma_2})=\frac{d}{dt}\left( \rho_{7,7}+\rho_{5,5}+\rho_{3,3}+\rho_{1,1}\right)=\gamma_c \times \left\{-P_1^{\varsigma_2}\{1-f_R(\varepsilon_{s}^2)+\frac{\gamma}{\gamma_c}P^{\varsigma_1}_{0,1}\}+P_{0}^{\varsigma_2}\{f_R(\varepsilon_s^2)+\frac{\gamma}{\gamma_c}P_{1,1}^{\varsigma_1}\}\right\},
	\label{eq:second_sys}
	\end{align}
\end{widetext}
\normalsize
	where  $\tau_{{4},{1}}$ and $\tau_{{1},{4}}$ are assumed to be zero and $\gamma=\tau_{{6},{3}}=\tau_{{3},{6}}$.  An intuitive approach to derive the QME for an arbitrary array with higher number of  Coulomb coupled quantum dots is detailed in the Supplementary Sec. \color{black}
	The sets of Eqns.~\eqref{eq:first_sys} and \eqref{eq:second_sys}  coupled to each other. To calculate the values of the state probabilities, these sets of equations may be solved numerically  using any iterative method. On solution of the state probabilities given by Eqns.~\eqref{eq:first_sys} and \eqref{eq:second_sys}, the charge current $I_{L(R)}$ through the system    can be calculated using the equations:
	\begin{align}
	I_L= & q\gamma_c \times \left\{P^{\varsigma_1}_{0,0}f_L(\varepsilon_s^1)+P^{\varsigma_1}_{0,1}f_L(\varepsilon_s^1+U)\right\}  - q\gamma_c P^{\varsigma_1}_{1,0}\{1-f_L(\varepsilon_s^1)\}- q\gamma_c P^{\varsigma_s^1}_{1,1}\{1-f_L(\varepsilon_s^1+U)\} \nonumber \\
	I_R= & -q\gamma_c \times \left\{P^{\varsigma_2}_{0}f_R(\varepsilon_s^1)-P^{\varsigma_2}_{1}\{1-f_R(\varepsilon_s^1)\}\right\} 
	\label{eq:final}
	\end{align}
	Next, I use the set of Eqns.~\eqref{eq:first_sys}-\eqref{eq:final}, to characterize the set-up demonstrated in  Fig.~\ref{fig:Fig1}. Without loss of generality, I assume that $\gamma_c=5\times 10^{-7}\frac{q}{h}$ and $ \gamma=5\times 10^{-6}\frac{q}{h}$. In particular, I show the  characteristics of the set-up, as captured by the proposed QME for three different cases: (i) fixed voltage bias, (ii) fixed temperature bias and (iii) varying voltage bias and capacitive coupling energy. Fig.~\ref{fig:Fig2}(a) demonstrates the regime of current flow through the system at $T_R=T_L=T_G=T=5K,~U=2meV ~(\approx 4.63\frac{kT}{q})$, and a voltage bias $V=1mV ~(\approx ~2.3\frac{kT}{q})$ for a range of   positions of $\varepsilon_g$ and $\varepsilon_s^1$. We note that the maximum current flow occurs when $\varepsilon_g$ goes a few $kT$ below the equilibrium Fermi energy $\mu_0$, that is when the level $\varepsilon_g$ is always occupied with an electron, as expected. Similarly, the current flow occurs when $\varepsilon_s^1+U-\mu_0$ lies in the bias   window, that is from $-qV/2<\varepsilon_s^1+U-\mu_0<qV/2$. Fig.~\ref{fig:Fig2}(b) demonstrates the regime of short-circuited thermoelectric current flow through the system for a temperature bias given by $T_R=T_L=5K,~T_G=10K$ and $~U=2meV ~(\approx 3.1\frac{kT}{q})$  for a range of position of $\varepsilon_g$ and $\varepsilon_s^1$. This short-circuited current flows by absorbing heat energy from  the reservoir $G$ and constitutes the non-local thermoelectric action proposed in the Refs.  \cite{coulomb_TE1,coulomb_TE2,coulomb_TE3,coulomb_TE4,coulomb_TE5,coulomb_TE6}. Fig.~\ref{fig:Fig2}(c) shows the regime of current flow (absolute value) with variation of the Coulomb coupling energy $U$ and the voltage bias $V$ at $T_R=T_L=T_G=T=5K$, $\varepsilon_s^1-\mu_0=-1.5meV~(\approx -3.5\frac{kT}{q})$ and $\varepsilon_g-\mu_0=-2meV~(\approx -4.63\frac{kT}{q})$. As expected, the current  magnitude increases to saturation with increase in magnitude of the applied bias $V$ and decreases with the increase in $U$ (since the electron occupancy in probability in $\varepsilon_g$ decreases with increase in $U$. In addition the energy level $\varepsilon_s^1+U$ moves outside the bias window with an increase in $U$).\\
	\indent To conclude, in this paper, I have methodically derived the QME  for a Coulomb coupled system with three quantum dots. The proposed QME has been derived from the system Hamiltonian using density matrix formulation and captures the intuitive aspects of the sequential electron transport and current flow. The framework elaborated in this paper can  be further extended  to derive QME of systems with  higher number of Coulomb coupled quantum dots. Hence, the formulation developed in this paper can pave the way towards an intuitive analysis of transport physics for an array of Coulomb coupled quantum dots in the sequential tunneling regime.\\
\appendix
\begin{widetext}
\section{Supplementary information}
\indent Here, I show an intuitive approach to write the quantum master equation (QME) for an arbitrary array of Coulomb coupled quantum dots.  I demonstrate two different arrangements and derive the quantum master equations (QME) from an intuitive perspective. Although these equations can also be mathematically derived from density matrix formulation, I stress on the fact that an understanding of the intuitive approach to write the QME for an arbitrary array of Coulomb coupled quantum dots circumvents clumsy mathematical derivations and is beneficial to study the behaviour of arbitrary Coulomb coupled systems. \\
\indent  I demonstrate an intuitive approach to derive the quantum master equations for an array with arbitrary pairs of  Coulomb coupled quantum dots with staircase ground state configuration. The two systems to be discussed in this context are demonstrated in Fig.~\ref{fig:supplementary}. Although the QME of such systems can be mathematically derived from density matrix formulation, I  elaborate the intuitive approach to write the system QME. Let us consider the system-I demonstrated in Fig.~\ref{fig:supplementary}(a). In this case, the top array of $N$ quantum dots $S_n$ share a staircase ground state  configuration with $\varepsilon_s^{j+1}=\varepsilon_s^{j}+U_m$.  The  dot $S_j$ is capacitively connected to the  dot $G_j$ with mutual charging energy $U_m$. The dots $G_j$  are, inturn,  electrically connected to the reservoir $G$.  Due to such staircase ground state configuration, in the limit of weak coupling and not too low value of $U_m$, we can safely assume that interdot tunneling between $S_{j-1}$ and $S_j$ can only occur when the round state of $G_{j-1}$ is occupied. $S_1$ and $S_N$ are connected to the reservoirs $L$ and $R$ respectively, while $S_j$ is electrically connected to the dots $S_{j-1}$ and $S_{j+1}$ for $2\leq j \leq N-1$.\\
\begin{figure*}
	\includegraphics[width=\textwidth]{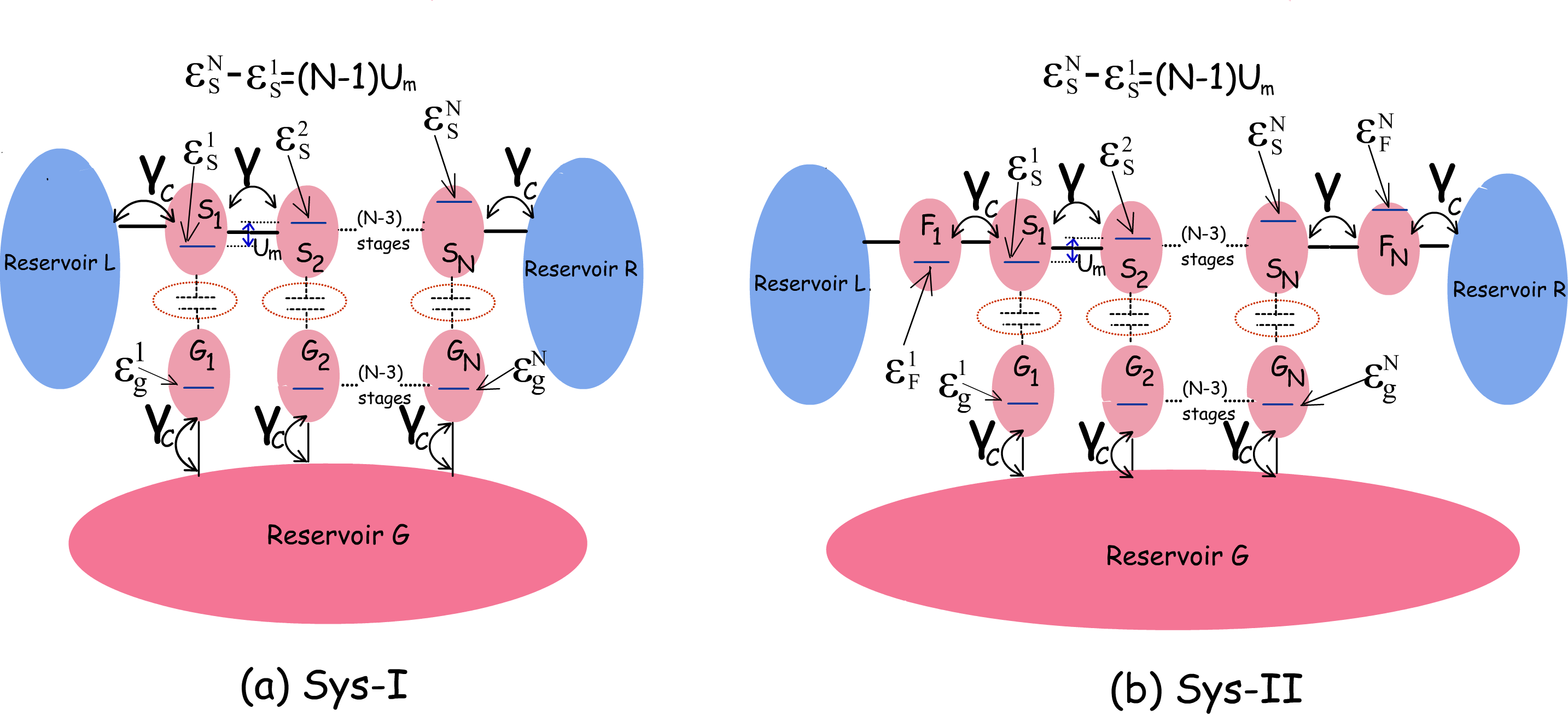}
	\caption{Schematic diagram illustrating two Coulomb coupled systems with arbitrary number of capacitively coupled quantum dots. (a) system-I: array with $N$ pair of Coulomb coupled dots, and  (b)  system-II: array with $N$ pair of capacitively coupled quantum dots with added filters $F_1$ and $F_N$ at the contact to dot interfaces. The capacitively coupled quantum dots share a staircase ground state configuration with $\varepsilon_{s}^j=\varepsilon_{s}^{j-1}+U_m$. The dots $F_1$ and $F_N$ in system-II donot share capacitive coupling with any other dot in the system. The ground state of $F_1$ and  $F_N$ are given by $\varepsilon_F^1=\varepsilon_{s}^1$ and $\varepsilon_F^N=\varepsilon_{s}^N+U_m$. }
	\label{fig:supplementary}
\end{figure*}
\indent In this case, following the previous convention, the entire system can be divided into sub-systems $\varsigma_n$, with $1\leq n\leq N$, $N$ being the total number of pairs of Coulomb coupled quantum-dots. Each sub-system $\varsigma_n$ consists of the pair of Coulomb coupled dots $S_n$ and $G_n$, with mutual charging energy $U_m$.  Following the same convention and assumptions as elaborated in the main text, I write the probability of occupancy of each subsystem $\varsigma_n$ as $P_{x,y}^{\varsigma_n}$, where $x$ and $y$ denote the number of electrons in the dot $S_n$ and $G_n$ respectively (Fig.~\ref{fig:supplementary}.a). Now, let us consider the quantity $\frac{d}{dt}(P_{0,0}^{\varsigma_1})$. The system can exit the state $P_{0,0}^{\varsigma_1}$ under the following circumstances:
\begin{enumerate}
	\item An electron may tunnel into $S_1$ from reservoir $L$. This accounts for a term proportional to $P_{0,0}^{\varsigma_1}f_L(\varepsilon_s^1)$. The sub-system now enters the state $P_{1,0}^{\varsigma_1}\{f_L(\varepsilon_s^1)$
	\item An electron may tunnel into the dot $G_1$ from $G$. This accounts for a term proportional to $P_{0,0}^{\varsigma_1}f_G(\varepsilon_g^1)$. The sub-system now enters the state $P_{0,1}^{\varsigma_1}$.
\end{enumerate}
Similarly, the system may enter into the state $P_{0,0}^{\varsigma_1}\{f_L(\varepsilon_s^1)$ from a different state. This happens in the following cases. 
\begin{enumerate}
	\item With the ground state of the dot $G_1$ being empty, an electron in $S_1$ tunnels out into $L$ and brings the sub-system from $P_{1,0}^{\varsigma_1}\{f_L(\varepsilon_s^1)$ to $P_{0,0}^{\varsigma_1}\{f_L(\varepsilon_s^1)$. This phenomenon can be accounted for by a term proportional to $P_{1,0}^{\varsigma_1}\{1-f_L(\varepsilon_s^1)\}$.
	\item With the ground state of the dot $S_1$ being unoccupied, an electron in $G_1$ tunnels out into $G$ and brings the sub-system from $P_{0,1}^{\varsigma_1}$ to $P_{0,0}^{\varsigma_1}$. This phenomenon can be accounted for by a term proportional to $P_{0,1}^{\varsigma_1}\{1-f_G(\varepsilon_g^1)\}$.
\end{enumerate} 
The sub-system rate equations for the quantity $P_{0,1}^{\varsigma_1}$ can thus be written as the sum of these four cases:

\begin{align}
\frac{d}{dt}(P_{0,0}^{\varsigma_1})=&\gamma_c \times \left\{-P_{0,0}^{\varsigma_1}\{f_L(\varepsilon_s^1)+f_G(\varepsilon_g^1)\}+P_{0,1}^{\varsigma_1}\{1-f_G(\varepsilon_g^1)\}+P_{1,0}^{\varsigma_1}\{1-f_L(\varepsilon_s^1)\}\right\}\nonumber 
\end{align}

Now let us consider the rate equation for $P_{1,1}^{\varsigma_1}$. The sub-system may exit from the state $P_{1,1}^{\varsigma_1}$ due to the following phenomena:
\begin{enumerate}
	\item The electron in $G_1$ may exit into the reservoir $G$ with energy $\varepsilon_{g}^1+U_m$ and the system may transit to the state $P_{1,0}^{\varsigma_1}$. This is can be taken into account by a term proportional to  $P_{1,1}^{\varsigma_1}[1-f_G(\varepsilon_{g}^1+U_m)]$.
	\item  The electron in $S_1$ may also exit into the reservoir $L$ with energy $\varepsilon_s^1+U_m$ and the system may transit to the state $P_{0,1}^{\varsigma_1}$. This is can be taken into account by a term proportional to  $P_{1,1}^{\varsigma_1}[1-f_L(\varepsilon_{s}^1+U_m)]$
	\item Finally, since the ground state of both the dots $S_1$ and $G_1$ are occupied, the electron in $S_1$ can tunnel into $S_2$, provided that the subsystem $\varsigma_2$ is in the state $P_{0,0}^{\varsigma_2}$. This is because the energy difference between the  ground states of $S_1$ and $S_2$ is $U_m$ and hence an electron in $S_1$ can only tunnel int $S_2$ when the ground states of both $S_1$ and $G_1$ are occupied, while the ground states of both $G_2$ and $S_2$ are empty. This phenomena can be taken into consideration via a term proportional to $P_{0,0}^{\varsigma_2}P^{\varsigma_1}_{1,1}$
\end{enumerate}
Similarly the sub-system may also transit into the state $P_{1,1}^{\varsigma_1}$  from other states. The phenomena responsible for the sub-system transit into the state $P_{1,1}^{\varsigma_1}$ include the following.
\begin{enumerate}
	\item With the ground state of $G_1$ already occupied, an electron may tunnel from $L$ into $S_1$ with an energy $\varepsilon_{s}^1+U_m$. Such tunneling transfers the system from $P_{0,1}^{\varsigma_1}$ to $P_{1,1}^{\varsigma_1}$. Such an event can be taken into account by a term proportional to $P_{0,1}^{\varsigma_1}f_L(\varepsilon_s^1+U_m)$.
	\item With the ground state of $S_1$ already occupied, an electron may tunnel into $G_1$ from $G$ at an energy $\varepsilon_g^1+U_m$. Such tunneling takes the system from $P_{1,0}^{\varsigma_1}$ to $P_{1,1}^{\varsigma_1}$. Such an event can be taken into account by a term proportional to $P_{1,0}^{\varsigma_1}f_G(\varepsilon_g^1+U_m)$.
	\item An electron can also tunnel from $S_2$ to $S_1$, provided that the ground state of $G_1$ and $S_2$ are occupied and the ground state of $G_2$ is empty. This process takes the sub-system from $P_{0,1}^{\varsigma_1}$ to $P_{1,1}^{\varsigma_1}$ and can be accounted for by a term proportional to $P_{0,1}^{\varsigma_1}P^{\varsigma_2}_{1,0}$.
\end{enumerate}
This  equation governing the sub-system state probability $P^{\varsigma_2}_{1,1}$ can thus be written as: 
\begin{align}
\frac{d}{dt}(P_{1,1}^{\varsigma_1})=&\gamma_c \times \left\{-P_{1,1}^{\varsigma_1}\left\{[1-f_G(\varepsilon_{g}^1+U_m)]+[1-f_L(\varepsilon_s^1+U_m)]+\frac{\gamma}{\gamma_C}P^{\varsigma_2}_{0,0}\right\}\right\} \nonumber \\ &+\gamma_c \left\{P_{1,0}^{\varsigma_1}f_G(\varepsilon_g^1+U_m) +P_{0,1}^{\varsigma_1}\left\{f_L(\varepsilon_s^1+U_m)+\frac{\gamma}{\gamma_c}P^{\varsigma_2}_{1,0}\right\}\right\}
\end{align}
The rate equations for $P_{0,1}^{\varsigma_1}$ and $P_{1,0}^{\varsigma_1}$ can be derived in a similar fashion Thus, the equations governing the sub-system state probabilities can be written as:
\begin{align}
\frac{d}{dt}(P_{0,0}^{\varsigma_1})=&\gamma_c \times \left\{-P_{0,0}^{\varsigma_1}\{f_L(\varepsilon_s^1)+f_G(\varepsilon_g^1)\}+P_{0,1}^{\varsigma_1}\{1-f_G(\varepsilon_g^1)\}+P_{1,0}^{\varsigma_1}\{1-f_L(\varepsilon_s^1)\}\right\}\nonumber \\
\frac{d}{dt}(P_{1,0}^{\varsigma_1})=&\gamma_c \times \left\{-P_{1,0}^{\varsigma_1}\left\{1-f_L(\varepsilon_{s}^1)+f_G(\varepsilon_g^1+U_m)\right\}+P_{1,1}^{\varsigma_1}\left\{1-f_G(\varepsilon_g^1+U_m)\right\}+P_{0,0}^{\varsigma_1}f_L(\varepsilon_s^1)\right\} \nonumber \\
\frac{d}{dt}(P_{0,1}^{\varsigma_1})=&\gamma_c \times \left\{-P_{0,1}^{\varsigma_1}\left\{1-f_G(\varepsilon_{g}^1)+f_L(\varepsilon_s^1+U_m)+\frac{\gamma}{\gamma_c}P^{\varsigma_2}_{1,0}\right\}\right\} \nonumber \\& +\gamma_c \left\{P_{0,0}^{\varsigma_1}f_G(\varepsilon_g^1)+P_{1,1}^{\varsigma_1}\left\{1-f_L(\varepsilon_s^1+U_m)+\frac{\gamma}{\gamma_c}P^{\varsigma_2}_{0,0}\right\}\right\} \nonumber \\
\frac{d}{dt}(P_{1,1}^{\varsigma_1})=&\gamma_c \times \left\{-P_{1,1}^{\varsigma_1}\left\{[1-f_G(\varepsilon_{g}^1+U_m)]+[1-f_L(\varepsilon_s^1+U_m)]+\frac{\gamma}{\gamma_C}P^{\varsigma_2}_{0,0}\right\}\right\} \nonumber \\ &+\gamma_c \left\{P_{1,0}^{\varsigma_1}f_G(\varepsilon_g^1+U_m) +P_{0,1}^{\varsigma_1}\left\{f_L(\varepsilon_s^1+U_m)+\frac{\gamma}{\gamma_c}P^{\varsigma_2}_{1,0}\right\}\right\} \nonumber \\
\label{eq:first_sys_1}
\end{align} 
Similarly, the rate equations governing the state probabilities  for the sub-system $N$ can be written as:
\begin{align}
\frac{d}{dt}(P_{0,0}^{\varsigma_N})=&\gamma_c \times \left\{-P_{0,0}^{\varsigma_N}\left\{f_R(\varepsilon_s^N)+f_G(\varepsilon_g^N)+\frac{\gamma}{\gamma_c}P_{1,1}^{\varsigma_{N-1}}\right\}+P_{0,1}^{\varsigma_N}\{1-f_G(\varepsilon_g^N)\}+P_{1,0}^{\varsigma_N}\left\{1-f_R(\varepsilon_s^N)+\frac{\gamma}{\gamma_c}P_{0,1}^{\varsigma_{N-1}}\right\}\right\}\nonumber \\
\frac{d}{dt}(P_{1,0}^{\varsigma_N})=&\gamma_c \times \left\{-P_{1,0}^{\varsigma_N}\left\{1-f_R(\varepsilon_{s}^N)+f_G(\varepsilon_g^N+U_m)+\frac{\gamma}{\gamma_c}P_{0,1}^{\varsigma_{N-1}}\right\}+P_{1,1}^{\varsigma_N}\left\{1-f_G(\varepsilon_g^N+U_m)\right\}+P_{0,0}^{\varsigma_N}\left\{f_R(\varepsilon_g^N)+\frac{\gamma}{\gamma_c}P_{1,1}^{\varsigma_{N-1}}\right\}\right\} \nonumber \\
\frac{d}{dt}(P_{0,1}^{\varsigma_N})=&\gamma_c \times \left\{-P_{0,1}^{\varsigma_N}\left\{1-f_G(\varepsilon_{g}^N)+f_R(\varepsilon_s^N+U_m)\right\}+P_{0,0}^{\varsigma_N}f_G(\varepsilon_g^N)+P_{1,1}^{\varsigma_N}\left\{1-f_R(\varepsilon_s^N+U_m)\right\}\right\} \nonumber \\
\frac{d}{dt}(P_{1,1}^{\varsigma_N})=&\gamma_c \times \left\{-P_{1,1}^{\varsigma_N}\left\{[1-f_G(\varepsilon_{g}^N+U_m)]+[1-f_R(\varepsilon_s^N+U_m)]\right\}+P_{1,0}^{\varsigma_N}f_G(\varepsilon_g^N+U_m) +P_{0,1}^{\varsigma_N}\left\{f_R(\varepsilon_s^N+U_m)\right\}\right\} \nonumber \\
\label{eq:last_sys_1}
\end{align} 
For the sub-systems $\varsigma_n$ for $2\leq n \leq (N-1)$, the rate equations are slightly different, since the these sub-systems are not connected to any reservoir. Let us consider the state probability $P_{0,0}^{\varsigma_n}$. A sub-system transition from the state $P_{0,0}^{\varsigma_n}$ to another state can occur due  to the following circumstances:
\begin{enumerate}
	\item An electron from $S_{n-1}$ can tunnel into $S_n$, provided that the ground states of $S_{n-1}$ and $G_{n-1}$ are occupied and the ground states of $S_n$ and  $G_n$ are empty. Such tunneling results in sub-system transition from $P_{0,0}^{\varsigma_n}$ to $P_{1,0}^{\varsigma_n}$. Such a process can be accounted in the rate equation via a term proportional to  $P_{0,0}^{\varsigma_n}P_{1,1}^{\varsigma_{n-1}}$.
	\item An electron can tunnel into $G_n$ from the reservoir $G$. Such process causes the  sub-system to transit from $P_{0,0}^{\varsigma_n}$ to $P_{0,1}^{\varsigma_n}$ and is proportional to $P_{0,0}^{\varsigma_n}f_G(\varsigma_g^n)$.
	
\end{enumerate}
Similarly the sub-system can transit into $P_{0,0}^{\varsigma_n}$ due to the following phenomena.
\begin{enumerate}
	\item Provided that the ground state of $S_n$ is empty, an electron present in the ground state of $G_n$ can tunnel out into reservoir $G$. Such tunneling results in subsystem transition from $P_{0,1}^{\varsigma_n}$ to $P_{0,0}^{\varsigma_n}$ and can be accounted by a term proportional to $P_{0,0}^{\varsigma_n}\{1-f_G(\varepsilon_g^n)\}$
	\item Provided that the ground states of $S_n$ and $G_{n-1}$ are occupied and that of $G_n$ and $S_{n-1}$ are empty, an electron can tunnel from $S_n$ to $S_{n-1}$ resulting in sub-system transition from $P_{1,0}^{\varsigma_n}$ to $P_{0,0}^{\varsigma_n}$. This phenomenon can be accounted by a term proportional to $P_{1,0}^{\varsigma_n}P_{0,1}^{\varsigma_{n-1}}$
\end{enumerate}
Thus, the rate equation governing $P_{0,0}^{\varsigma_n}$ for $2\leq n \leq N-1$ can be given by: 
\begin{align}
\frac{d}{dt}(P_{0,0}^{\varsigma_n})=&\gamma_c \times \left\{-P_{0,0}^{\varsigma_n}\left\{\frac{\gamma}{\gamma_c}P_{1,1}^{\varsigma_{n-1}}+f_G(\varepsilon_g^n)\right\}+P_{0,1}^{\varsigma_n}\{1-f_G(\varepsilon_g^n)\}+\frac{\gamma}{\gamma_c}P_{1,0}^{\varsigma_n}P_{0,1}^{\varsigma_{n-1}}\right\}
\end{align}
In a similar way, the rate equations governing the various sub-system probabilities, for $2\leq n \leq N-1$,  can be written as:
\begin{align}
\frac{d}{dt}(P_{0,0}^{\varsigma_n})=&\gamma_c \times \left\{-P_{0,0}^{\varsigma_n}\left\{\frac{\gamma}{\gamma_c}P_{1,1}^{\varsigma_{n-1}}+f_G(\varepsilon_g^n)\right\}+P_{0,1}^{\varsigma_n}\{1-f_G(\varepsilon_g^n)\}+\frac{\gamma}{\gamma_c}P_{1,0}^{\varsigma_n}P_{0,1}^{\varsigma_{n-1}}\right\}\nonumber \\
\frac{d}{dt}(P_{1,0}^{\varsigma_n})=&\gamma_c \times \left\{-P_{1,0}^{\varsigma_n}\left\{\frac{\gamma}{\gamma_c}P_{0,1}^{\varsigma_{n-1}}+f_G(\varepsilon_g^n+U_m)\right\}+P_{1,1}^{\varsigma_n}\left\{1-f_G(\varepsilon_g^n+U_m)\right\}+\frac{\gamma}{\gamma_c}P_{0,0}^{\varsigma_n}P_{1,1}^{\varsigma_{n-1}}\right\} \nonumber \\
\frac{d}{dt}(P_{0,1}^{\varsigma_n})=&\gamma_c \times \left\{-P_{0,1}^{\varsigma_n}\left\{1-f_G(\varepsilon_{g}^n)+\frac{\gamma}{\gamma_c}P^{\varsigma_{n+1}}_{1,0}\right\}+P_{0,0}^{\varsigma_n}f_G(\varepsilon_g^n)+\frac{\gamma}{\gamma_c}P_{1,1}^{\varsigma_n}P^{\varsigma_{n+1}}_{0,0}\right\} \nonumber \\
\frac{d}{dt}(P_{1,1}^{\varsigma_n})=&\gamma_c \times \left\{-P_{1,1}^{\varsigma_n}\left\{[1-f_G(\varepsilon_{g}^n+U_m)]+\frac{\gamma}{\gamma_C}P^{\varsigma_{n+1}}_{0,0}\right\}+P_{1,0}^{\varsigma_n}f_G(\varepsilon_g^n+U_m) +\frac{\gamma}{\gamma_c}P_{0,1}^{\varsigma_n}P^{\varsigma_{n+1}}_{1,0}\right\} \nonumber \\
\label{eq:middle_sys}
\end{align} 
The set of Eqns. \eqref{eq:first_sys_1}, \eqref{eq:last_sys_1} and \eqref{eq:middle_sys} constitute the QME for the system shown in Fig. \ref{fig:supplementary}(a). As discussed in the main text, these sets of Eqns. are coupled to each other and  can be solved using any iterative numerical techniques. On solution of the state probabilities, the various electrical properties of the system can be determined.\\ 
\indent 	Next, let us consider the system shown in \ref{fig:supplementary}(b). In this figure two energy quantum dots $F_1$ and $F_2$ acting as energy filters are added between the interface of $L-S_1$ and $R-S_N$. The dots $F_1$ and $F_N$ are not Coulomb coupled to any other dot in the system. In the same way as before, we divide the entire system into sub-systems. The dots $F_1$ and $F_N$ constitute sub-systems $\varsigma_0$ and $\varsigma_{N+1}$, while the combination of the dots  $S_j$ and $G_j$ constitute the sub-system $\varsigma_j$ (for $1\leq j \leq N$). In what follows, $P_z^{\varsigma_{0(N+1)}}$ will be used to denote the state probability of the sub-systems $\varsigma_{0(N+1)}$, with $z$ denoting the number of electrons in the dot $F_{1(N)}$. $P_{x,y}^{\varsigma_j}$, on the other hand, will be used to denote the state probability of the sub-system $j$ ($j\neq 0,~ N+1$), where $x$ and $y$ denote the number of electrons in the dots $S_j$ and $G_j$ respectively. The ground state configurations of the filter dots $F_1$ and $F_N$ are given by $\varepsilon_F^1=\varepsilon_s^1$ and $\varepsilon_F^N=\varepsilon_s^N+U_m$. Such arrangement of quantum dots with energy filters have been suggested to enhance non-local waste heat harvesting in heat engines based on Coulomb coupled systems \cite{nonlocal}.  Just like the previous approach, we can intuitively write the rate equations for the sub-systems as follows.\\
\indent \textbf{Rate equations for the sub-systems $\varsigma_0$ and $\varsigma_{N+1}$}
\begin{align}
\frac{d}{dt}(P_{0}^{\varsigma_0})=&\gamma_c \times \left\{-P_{0}^{\varsigma_0}f_L(\varepsilon_F^1)+P_{1}^{\varsigma_0}\{1-f_L(\varepsilon_F^1)\}+\frac{\gamma}{\gamma_c}\left\{- P_{0}^{\varsigma_0} P_{1,0}^{\varsigma_1}+P_{1}^{\varsigma_0} P_{0,0}^{\varsigma_1}\right\}\right\}\nonumber \\
\frac{d}{dt}(P_{1}^{\varsigma_0})=&\gamma_c \times \left\{P_{0}^{\varsigma_0}f_L(\varepsilon_F^1)-P_{1}^{\varsigma_0}\{1-f_L(\varepsilon_F^1)\}+\frac{\gamma}{\gamma_c}\left\{ P_{0}^{\varsigma_0} P_{1,0}^{\varsigma_1}-P_{1}^{\varsigma_0} P_{0,0}^{\varsigma_1}\right\}\right\}\nonumber \\
\frac{d}{dt}(P_{0}^{\varsigma_{N+1}})=&\gamma_c \times \left\{-P_{0}^{\varsigma_{N+1}}f_R(\varepsilon_F^N)+P_{1}^{\varsigma_{N+1}}\{1-f_R(\varepsilon_F^N)\}+\frac{\gamma}{\gamma_c}\left\{- P_{0}^{\varsigma_{N+1}} P_{1,1}^{\varsigma_N}+P_{1}^{\varsigma_{N+1}} P_{0,1}^{\varsigma_N}\right\}\right\}\nonumber \\
\frac{d}{dt}(P_{1}^{\varsigma_{N+1}})=&\gamma_c \times \left\{P_{0}^{\varsigma_{N+1}}f_R(\varepsilon_F^N)-P_{1}^{\varsigma_{N+1}}\{1-f_R(\varepsilon_F^N)\}+\frac{\gamma}{\gamma_c}\left\{ P_{0}^{\varsigma_{N+1}} P_{1,1}^{\varsigma_N}-P_{1}^{\varsigma_{N+1}} P_{0,1}^{\varsigma_N}\right\}\right\}
\label{eq:first_sys_11}
\end{align} 

\indent \textbf{Rate equations for the sub-systems $\varsigma_1$ and $\varsigma_{N}$}
\begin{align}
\frac{d}{dt}(P_{0,0}^{\varsigma_1})=&\gamma_c \times \left\{-P_{0,0}^{\varsigma_1}\left\{\frac{\gamma}{\gamma_c}P_{1}^{\varsigma_0}+f_G(\varepsilon_g^1)\right\}+P_{0,1}^{\varsigma_1}\{1-f_G(\varepsilon_g^1)\}+\frac{\gamma}{\gamma_c}P_{1,0}^{\varsigma_1}P_{0}^{\varsigma_0}\right\}\nonumber \\
\frac{d}{dt}(P_{1,0}^{\varsigma_1})=&\gamma_c \times \left\{-P_{1,0}^{\varsigma_1}\left\{\frac{\gamma}{\gamma_c}P_{0}^{\varsigma_0}+f_G(\varepsilon_g^1+U_m)\right\}+P_{1,1}^{\varsigma_1}\left\{1-f_G(\varepsilon_g^1+U_m)\right\}+\frac{\gamma}{\gamma_c}P_{0,0}^{\varsigma_1}P_{1}^{\varsigma_0}\right\} \nonumber \\
\frac{d}{dt}(P_{0,1}^{\varsigma_1})=&\gamma_c \times \left\{-P_{0,1}^{\varsigma_1}\left\{1-f_G(\varepsilon_{g}^1)+\frac{\gamma}{\gamma_c}P^{\varsigma_2}_{1,0}\right\}+P_{0,0}^{\varsigma_1}f_G(\varepsilon_g^1)+\frac{\gamma}{\gamma_c}P_{1,1}^{\varsigma_1}P^{\varsigma_2}_{0,0}\right\} \nonumber \\
\frac{d}{dt}(P_{1,1}^{\varsigma_1})=&\gamma_c \times \left\{-P_{1,1}^{\varsigma_1}\left\{[1-f_G(\varepsilon_{g}^1+U_m)]+\frac{\gamma}{\gamma_C}P^{\varsigma_2}_{0,0}\right\}+P_{1,0}^{\varsigma_1}f_G(\varepsilon_g^1+U_m) +\frac{\gamma}{\gamma_c}P_{0,1}^{\varsigma_1}P^{\varsigma_2}_{1,0}\right\} 
\end{align}

\begin{align}
\frac{d}{dt}(P_{0,0}^{\varsigma_N})=&\gamma_c \times \left\{-P_{0,0}^{\varsigma_N}\left\{f_G(\varepsilon_g^N)+\frac{\gamma}{\gamma_c}P_{1,1}^{\varsigma_{N-1}}\right\}+P_{0,1}^{\varsigma_N}\{1-f_G(\varepsilon_g^N)\}+\frac{\gamma}{\gamma_c}P_{1,0}^{\varsigma_N}P_{0,1}^{\varsigma_{N-1}}\right\}\nonumber \\
\frac{d}{dt}(P_{1,0}^{\varsigma_N})=&\gamma_c \times \left\{-P_{1,0}^{\varsigma_N}\left\{f_G(\varepsilon_g^N+U_m)+\frac{\gamma}{\gamma_c}P_{0,1}^{\varsigma_{N-1}}\right\}+P_{1,1}^{\varsigma_N}\left\{1-f_G(\varepsilon_g^N+U_m)\right\}+\frac{\gamma}{\gamma_c}P_{0,0}^{\varsigma_N}P_{1,1}^{\varsigma_{N-1}}\right\} \nonumber \\
\frac{d}{dt}(P_{0,1}^{\varsigma_N})=&\gamma_c \times \left\{-P_{0,1}^{\varsigma_N}\left\{1-f_G(\varepsilon_{g}^N)+\frac{\gamma}{\gamma_c}P_1^{\varsigma_{N+1}}\right\}+P_{0,0}^{\varsigma_N}f_G(\varepsilon_g^N)+\frac{\gamma}{\gamma_c}P_{1,1}^{\varsigma_N}P_0^{\varsigma_{N+1}}\right\} \nonumber \\
\frac{d}{dt}(P_{1,1}^{\varsigma_N})=&\gamma_c \times \left\{-P_{1,1}^{\varsigma_N}\left\{[1-f_G(\varepsilon_{g}^N+U_m)]+\frac{\gamma}{\gamma_c}P_0^{\varsigma_{N+1}}\right\}+P_{1,0}^{\varsigma_N}f_G(\varepsilon_g^N+U_m) +\frac{\gamma}{\gamma_c}P_1^{\varsigma_{N+1}}P_{0,1}^{\varsigma_N}\right\} \nonumber \\
\label{eq:last_sys_18}
\end{align} 
\indent \textbf{Rate equations for the sub-systems $\varsigma_n$ for $2\leq n \leq N-1$}	
\begin{align}
\frac{d}{dt}(P_{0,0}^{\varsigma_n})=&\gamma_c \times \left\{-P_{0,0}^{\varsigma_n}\left\{\frac{\gamma}{\gamma_c}P_{1,1}^{\varsigma_{n-1}}+f_G(\varepsilon_g^n)\right\}+P_{0,1}^{\varsigma_n}\{1-f_G(\varepsilon_g^n)\}+\frac{\gamma}{\gamma_c}P_{1,0}^{\varsigma_n}P_{0,1}^{\varsigma_{n-1}}\right\}\nonumber \\
\frac{d}{dt}(P_{1,0}^{\varsigma_n})=&\gamma_c \times \left\{-P_{1,0}^{\varsigma_n}\left\{\frac{\gamma}{\gamma_c}P_{0,1}^{\varsigma_{n-1}}+f_G(\varepsilon_g^n+U_m)\right\}+P_{1,1}^{\varsigma_n}\left\{1-f_G(\varepsilon_g^n+U_m)\right\}+\frac{\gamma}{\gamma_c}P_{0,0}^{\varsigma_n}P_{1,1}^{\varsigma_{n-1}}\right\} \nonumber \\
\frac{d}{dt}(P_{0,1}^{\varsigma_n})=&\gamma_c \times \left\{-P_{0,1}^{\varsigma_n}\left\{1-f_G(\varepsilon_{g}^n)+\frac{\gamma}{\gamma_c}P^{\varsigma_{n+1}}_{1,0}\right\}+P_{0,0}^{\varsigma_n}f_G(\varepsilon_g^n)+\frac{\gamma}{\gamma_c}P_{1,1}^{\varsigma_n}P^{\varsigma_{n+1}}_{0,0}\right\} \nonumber \\
\frac{d}{dt}(P_{1,1}^{\varsigma_n})=&\gamma_c \times \left\{-P_{1,1}^{\varsigma_n}\left\{[1-f_G(\varepsilon_{g}^n+U_m)]+\frac{\gamma}{\gamma_C}P^{\varsigma_{n+1}}_{0,0}\right\}+P_{1,0}^{\varsigma_n}f_G(\varepsilon_g^n+U_m) +\frac{\gamma}{\gamma_c}P_{0,1}^{\varsigma_n}P^{\varsigma_{n+1}}_{1,0}\right\} \nonumber \\
\label{eq:middle_sys1}
\end{align} 	
Like the previous case, the set of Eqns. \eqref{eq:first_sys_11}-\eqref{eq:middle_sys1} constitute the entire set of QME of the system demonstrated in Fig.~\ref{fig:supplementary}(b). The set of equations are coupled to each other and can be solved using any iterative numerical scheme.\\
\end{widetext}
 \bibliography{apssamp}
\end{document}